\documentclass[aps,pre,%
twocolumn,%
amssymb,amsmath,%
showpacs,showkeys,%
tightenings,%
ejectum,%
breveted]{revtex4}
\usepackage[final]{graphicx}
\usepackage{dcolumn}
\begin{document}
%--------------------------------------
\title{Asymptotic Properties of Path Integral Ideals}
\author{A. Bogojevi\'c}
\author{A. Bala\v{z}}
\author{A. Beli\'c}
\affiliation{Institute of Physics, P.O. Box 57, 11001
Belgrade, Serbia and Montenegro}
%--------------------------------------
%\date{\today}
%------------------------------------------------------------------------------
\begin{abstract}
We introduce and analyze a new quantity, the path integral ideal, governing the flow of generic discrete theories to the continuum limit and greatly increasing their convergence. The said flow is classified according to the degree of divergence of the potential at spatial infinity. Studying the asymptotic behavior
of path integral ideals we isolate the dominant terms in the effective potential that determine the behavior of a generic theory for large discrete time steps.
\end{abstract}
%------------------------------------------------------------------------------
\preprint{SCL preprint}
\maketitle
%------------------------------------------------------------------------------

%------------------------------------------------------------------------------

Path integrals present a rich and flexible formalism for dealing with quantum and statistical theories \cite{feynmanhibbs,feynman} that has proven extremely useful for handling symmetries, deriving non-perturbative results, establishing connections between different theories \cite{itzyksonzuberplus}, and extending the quantization procedure to ever more complicated systems. They have served as catalysts for the exchange of key ideas between different areas of physics, most notably high energy and condensed matter physics \cite{itzyksondrouffeplus}. Today, analytical and numerical approaches to path integrals \cite{barker2,ceperley2} play important roles not only in physics but also in chemistry and materials science, and are acquiring a prominent role in mathematics and modern finance \cite{kleinert}.

Further development of the path integral method is constrained by the small number of solvable models, as well as by our rather limited knowledge of their precise mathematical properties. In fact, most of our knowledge is negative, e.g. we know which trajectories do not contribute to the path integral rather than which do. One of the few positive statements concerning path integrals is that relevant trajectories exhibit stochastic self-similarity \cite{feynmanhibbs}. As a result they have non-trivial fractal dimension and jaggedness \cite{kroegerplus}.
Researchers working on numerical approaches to path integrals have successfully utilized these kinematic consequences of self-similarity to produce efficient path-generating algorithms \cite{ceperley2}.

In a recent series of papers \cite{prlprb,pla05a} we have investigated the dynamical implications of stochastic self-similarity by
studying the relation between discretizations of path integrals with different coarseness. This has resulted in a systematic analytical procedure that may be used to reduce path integral error to $O(\epsilon^p)$ for arbitrary $p\in \mathbb{N}$, where $\epsilon$ is the discrete time step. Note that $\epsilon=T/N$, $T$ being the time of propagation and $N$ the discretization coarseness. This reduction of error brings about a substantial increase in the speed of numerical algorithms. Self-similarity played a crucial role in this procedure in that it allowed us to derive an integral equation relating discretized theories viewed at different coarseness and to solve it in terms of an asymptotic series. The asymptotic expansion, however, implies that the obtained method is directly applicable only for $\epsilon<1$.

The fact that we can arbitrarily decrease the error points to the possibility that one can extend the formalism and obtain exact information (i.e not given as a power series in $\epsilon$, and so valid even for large values of $\epsilon$) about the continuum theory. Large $\epsilon$ corresponds to long times of propagation, precisely what interests us in quantum field theory (or in modern finance). Equivalently, in condensed matter and materials science this corresponds to the physically most interesting region of small temperatures. Large $\epsilon$ behavior is also central for determining the energy spectrum of a given model, and as such is applicable in many areas of physics \cite{kleinert} (e.g. atomic and molecular physics, quantum dots).

In this paper we introduce and analyze a key new quantity, the path integral ideal, which governs the flow of generic discrete theories to the continuum. We show that this flow is classified according to the degree of divergence of the potential at spatial infinity.
We isolate the dominant part of the effective potential and show that it mediates the large $\epsilon$ behavior of the theory.
In this way we extend the previous method and derive the first properties valid for large $\epsilon$.

In \cite{pla05a} we showed that at the core of the new method one finds a generalization of Euler's summation formula to path integrals.
The Euler sum formula for the ordinary integral $\int_0^T dt f(t)$ can be viewed as an explicit construction of an ideal discretization
$f^*(t;\epsilon)$ of the integrand, where $\epsilon=T/N$ is the discrete time step, and $f^*(t;\epsilon)\to f(t)$ in the continuum limit $\epsilon\to 0$.
The discretization is ideal because for any coarseness $N$ it exactly reproduces the integral we seek to evaluate $\sum_{n=1}^N\epsilon\,f^*(t_n;\epsilon)=\int_0^T dt\, f(t)$, where $t_n=n\epsilon$.
The Euler sum formula gives $f^*(t;\epsilon)$ as an infinite power series in $\epsilon$.
The result is $f^*=f - \frac{\epsilon}{2} \dot f - \frac{2\epsilon^2}{3} \ddot f+\ldots$.
Truncation of $f^*$ to the first $p$ terms depends on $f(t)$ and its first $(p-1)$ derivatives and determines the integral up to $O(\epsilon^p)$.

The generalization of this to path integrals \cite{pla05a} showed the existence of an ideal discretized effective action $S^*$ having the property that
at any discretization coarseness $N$ its discretized amplitudes exactly reproduce the corresponding continuum amplitudes calculated with the starting action $S$.
For (Euclidean) actions of the form $S=\int_0^Tdt\,\left(\frac{1}{2}\, \dot q^2+V(q)\right)$,
the ideal discretized effective action was shown to be
\begin{equation}
S^*=\sum_{n=o}^{N-1}\left(\frac{\delta_n^2}{2\epsilon}+\epsilon\, W^*(\delta_n,\bar q_n; \epsilon)\right)\ ,
\end{equation}
where $\delta_n=q_{n+1}-q_n$ are the discretized velocities and $\bar q_n=\frac{1}{2}(q_{n+1}+q_n)$.
For obvious reasons we call the function $W^*(\delta,q;\epsilon)$ the path integral ideal.
We have shown \cite{pla05a} that this function is even in the first argument, that in the continuum limit it goes over to the potential $V(q)$,
and that it satisfies a novel integral equation relating discretizations of different coarseness:
\begin{widetext}
\begin{equation}
\label{integralequation}
e^{-\epsilon\, W^*(\delta,q; \epsilon)}= \sqrt{\frac{2}{\pi\epsilon}}\int_{-\infty}^{+\infty} dy\,
e^{-2y^2/\epsilon}
\,\exp\Bigg(-\frac{\epsilon}{2}\,W^*\left(\frac{\delta}{2}-y,q+\frac{\delta}{4}+\frac{y}{2}; \frac{\epsilon}{2}\right)
-\frac{\epsilon}{2}\,W^*\left(\frac{\delta}{2}+y,q-\frac{\delta}{4}+\frac{y}{2}; \frac{\epsilon}{2}\right)\Bigg)\ .
\end{equation}
\end{widetext}
In previous papers we asymptotically expanded the integral in the above equation obtaining a solution in the form of an infinite power series
in $\epsilon$, namely
\begin{equation}
\label{sum}
W^*(\delta,q;\epsilon)=\sum_{n=0}^\infty C_n(\delta,q)\,\epsilon^n\ .
\end{equation}
The coefficient functions $C_n(\delta,q)$ were found to be given in terms of algebraic equations.
Using MATHEMATICA 5.0 we obtained explicit solutions for the first nine coefficients.
The truncation of $W^*$ to the first $p$ terms depends on the potential $V$ and its first $(2p-2)$ derivatives and allows us to calculate the path integral
up to an $O(\epsilon^p)$ error term.
This has brought about a substantial speedup in the numerical calculations of path integrals.

Our aim is to study the asymptotic properties of the above sum.
Similar asymptotic expansions often arise in physics.
For example, perturbative expansion of path integrals (expansion in some small coupling constant $g$) generally leads to infinite sums of the form
$\sum_nA_ng^n$ where, for large $n$, the coefficients go as $A_n\sim (-)^nn!$.
Therefore, in analyzing the asymptotic properties of perturbative expansions one typically encounters sums like $\sum_n (-)^nn!g^n$.
Using Stirling's formula we find that the $n$-th term in the sum is asymptotically $(-gn)^n$.
Hence, the sum is divergent since its terms start increasing after $n_c\sim 1/g$.
The divergence is a result of a cavalier interchange of infinite sums and integrals at the very beginning of the asymptotic expansion procedure.
Although divergent, asymptotic sums give excellent approximations (for small $g$).
The best approximations are obtained by truncating the sum at $n_c$.
In principle one can often do much better than that. For example, leading strong coupling behavior may be obtained using Borel resummation \cite{itzyksondrouffeplus}.
More sophisticated techniques based on the variational perturbation method allow us even to extract sub-leading strong coupling behavior 
\cite{kleinertplus}.
This is the ultimate reason why we are interested in the asymptotic properties of path integral ideals, since, as it stands, the sum in eq.~(\ref{sum}) can only be used for small $\epsilon$. Exact, in this sense, means valid for all $\epsilon$.
\begin{figure}[!hb]
\includegraphics[width=8.5cm]{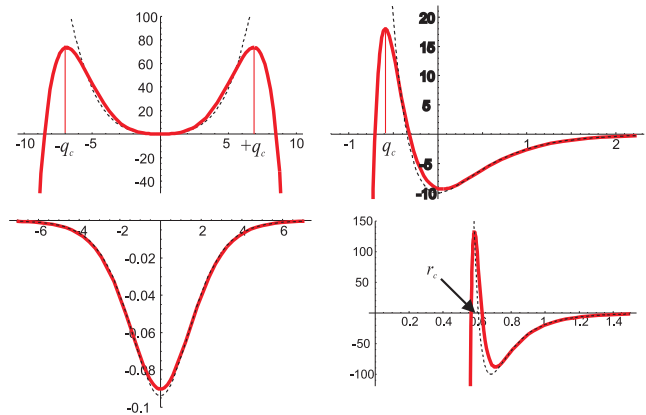}
\caption{(Color online) Effective potentials at level $p=3$ (solid thick line) and their critical points for the case of four different models:
oscillator with quartic coupling $V_1(q)=\frac{1}{2}q^2+\frac{1}{4!}g\,q^4$ (the case $\epsilon=0.5$, $g=1$ is shown top left);
particle moving in a Morse potential $V_2(q)=A(e^{-2\lambda q}-2e^{-\lambda q})$ (the case $\epsilon=0.1$, $A=10$, $\lambda=2$ is shown top right);
particle moving in a modified P\"oschl-Teller potential $V_3(q)=\frac{1}{2}\frac{\alpha^2\beta(\beta-1)}{\cosh^2(\alpha q)}$
(the case $\epsilon=1$, $\alpha=0.5$, $\beta=1.5$ is shown bottom left);
particle in a Lennard-Jones potential $V_4(r)=\frac{C}{r^{12}}-\frac{D}{r^6}$
(the case $\epsilon=0.01$, $C=1$, $D=20$ is shown bottom right). For comparison the original potentials have been plotted with a dashed thin line.}
\end{figure}

It is instructive to begin our investigation of the asymptotic properties of ideals by looking at the corresponding expressions at the $p=3$ level.
In fact, rather than look at the whole ideal we first focus on its velocity independent part -- the effective potential $V^*(q;\epsilon)=W^*(0,q;\epsilon)$.
At the $p=3$ level we found \cite{prlprb}
\begin{equation}
\label{p3}
V^*=V+\epsilon\frac{V''}{12}+\epsilon^2\left(-\frac{{V'}^2}{24}+\frac{V^{(4)}}{240}\right)\ .
\end{equation}
Note that, as a consequence of the ${V'}^2$ term being negative, it is possible for the above effective potential to be unbounded from bellow.
The top left plot in Fig.~1 shows the $p=3$ level effective potential for the case of an oscillator with quartic coupling
$V_1(q)=\frac{1}{2}q^2+\frac{1}{4!}g\,q^4$.
We see that the corresponding effective potential has critical points $\pm q_c$, where $q_c\sim\frac{1}{\epsilon\sqrt{g}}$.
Problems arise when using this effective action in the path integral to integrate over trajectories with $|q|>q_c$.
The top right plot in Fig.~1 shows the effective potential for a particle moving in a Morse potential $V_2(q)=A(e^{-2\lambda q}-2e^{-\lambda q})$.
In this case there is only one critical point $q_c\sim\frac{\ln\epsilon}{\lambda}$.
These two examples illustrate a general property: effective potentials
become unbounded from bellow at large positive (negative) values of $q$
if the original potential diverges at positive (negative) spatial infinity.
For example, the modified P\"oschl-Teller potential $V_3(q)=-\frac{\alpha^2\beta(\beta-1)}{2\cosh^2(\alpha q)}$
does not diverge at infinity and leads to a well behaved effective potential (bottom left plot in Fig.~1).
Another interesting class of potentials are those that diverge at a point.
From eq.~(\ref{p3}) we see that potentials that diverge at a point $q_0$ as $(q-q_0)^{-n}$ have no critical points for $n\le 2$,
the best example of this being the Coulomb potential.
The Lennard-Jones potential $V_4(r)=C/r^{12}-D/r^6$ diverges faster at the origin giving its $p=3$
effective potential a critical point (bottom right plot of Fig.~1).

Unboundedness from bellow is specific to certain $p$ levels only, e.g. $p=5$ level effective potentials are well behaved for all models.
At first hand it is surprising to have unboundedness from bellow for any model at any $p$ level.
To make sense of these effective potentials one needs to cut off coordinates at the critical points.
In fact, far from surprising, this is just the analogue of truncating asymptotic series to a critical number of terms $n_c$.

In solving an ordinary integral in an asymptotic expansion about a small parameter $g$ one obtains an infinite sum.
In many cases this sum is not convergent but asymptotic and the best approximation is obtained by truncating it at $n_c$.
As already shown $n_c$ typically behaves as $1/g$, or as some power of this, depending on the precise integral being evaluated.
In deriving the integral equation for ideals we asymptotically expanded path integrals
(the discrete time step $\epsilon$ playing the role of small parameter) and obtained the result in the form of an ordinary integral.
We have now come to the conclusion that in many cases this integral is not convergent but asymptotic
(depending on the behavior of the original potential at infinity) and that the best approximation is obtained when the effective potential
(and ideal) are truncated at $q_c$.
Typically $q_c$ behaves as $1/\epsilon$, or some power of this (depending on the model we are working with).

Let us point out that the original Monte Carlo simulations used to check the analytically derived results \cite{prlprb,pla05a}
worked well without any need for truncation at $q_c$.
The reason is that the path generating procedures in numerical simulations effectively create their own cut-offs.
The width of generated trajectories is typically $\sigma=\sqrt{\epsilon/2}$. The requirement that trajectories do not overshoot the critical point is
that $q_c$ is much larger than $\sigma$. For the oscillator with quartic coupling this is achieved for $N\gtrsim (g/2)^{1/3}T$.
For the case $g=10$ and $T=1$, for example, this criterion is satisfied even for the smallest $N$'s (roughest discretizations)
and in those simulations the critical points were never reached.

Asymptotic behavior of ideals is determined by the places where the original potential diverges as well as the degree of that divergence.
Finite potentials lead to well behaved effective potentials and ideals. An illustration of this is the modified P\"oschl-Teller potential.
Similarly, the Morse and Lennard-Jones potentials have path integral ideals that are well behaved at right infinity
(i.e. where the original potentials do not diverge).

In the remainder we will concentrate on potentials whose only divergence is at spatial infinity.
If the dominant behavior of $V$ at infinity is $q^n$ then from eq.~(\ref{p3}) we see that this is also the dominant behavior of the positive part of the
$p=3$ level effective potenital, while the negative part goes as $\epsilon^2 q^{2n-2}$.
The effective potential is well behaved for $n< 2$.
The case $n=2$ is integrable and so need not interest us here.
Still, let us note that it too is well behaved provided $\epsilon$ is small enough.
The case $n>2$ produces a critical point.
The simplest way to estimate the critical point is to note that in its vicinity the positive and negative terms in the effective potential balance out.
Therefore, for potentials that diverge with a power $n>2$, we have $q_c\sim1/\epsilon^{2/(n-2)}$.
Similarly, for potentials diverging exponentially we find $q_c\sim -\ln \epsilon^2$.

From studying the effective potential at the $p=3$ level we have learned which potentials give well behaved effective actions (and ideals)
and which have critical points, and have evaluated these critical points from the way the original potentials diverge at infinity.
We have also seen that it is important to study the ideal in the vicinity of the critical points, where the great majority of
its terms are subdominant.
It is not difficult to see that the dominant part of the complete $p=9$ level solution \cite{prlprb} equals
\begin{eqnarray}
\label{dominant}
&&V^*_{dom}=V-\epsilon^2\frac{{V'}^2}{24}+\epsilon^4\frac{{V'}^2}{240}V''\,-\nonumber\\
&&-\,\epsilon^6 \frac{{V'}^2}{6720}\left[V'V^{(3)}+\frac{17}{6}{V''}^2\right]+\\
&&+\,\epsilon^8\frac{{V'}^2}{241920}\left[{V'}^2V^{(4)}+11V'V''V^{(3)}+\frac{31}{3}{V''}^{(3)}\right]-\ldots\nonumber
\end{eqnarray}
We have written the dominant part of the ideal as $V^*_{dom}$ as it is velocity independent and part of the effective potential.
We have numerically checked that for all models considered this gives an excellent approximation to the ideal even far from critical points.
From the above formula we see that effective potentials are well behaved (bounded from bellow) for $p=1,2,5,6,9,10,\ldots$,
it is only at the remaining $p$ levels that one may find critical points.
This is the first indication that critical points are artifacts of the asymptotic expansion procedure and are not present in the exact path integral ideal.

The terms in the above expression can be determined directly (up to numerical factors) without first solving for the ideal.
To do this we introduce the following index: $\mbox{index} (ab) =  \mbox{index}(a) + \mbox{index}(b)$,
$\mbox{index} (V^{(\alpha)}) =  2+\alpha$, $\mbox{index} (\epsilon) =  \mbox{index}(\delta^2) = -2$.
Note also that the difference $W^*-V$ depends only on the derivatives of $V$.
It follows that $\mbox{index} (W^*)=2$, i.e all the terms in the ideal have the same index, as can be shown directly from eq.~(\ref{integralequation}).

To determine the dominant terms at spatial infinity we need to: (a) look at terms that are products of as many $V$'s as possible,
(b) have as few derivatives as possible, (c) pick out terms with smallest power of $\epsilon$.
All of this needs to be done with the constraint that the index remains fixed and equal to 2.
The fact that the dominant part of the ideal is velocity independent directly follows from the relation $\mbox{index}(\delta^2)=-\mbox{index}(V)$.
Every added power of $\delta^2$ (ideals are even in the velocities) removes one power of $V$ from the term and so decreases its order of divergence.
One can also show that leading terms proportional to $\epsilon^{2m+1}$ are subdominant to leading $\epsilon^{2m}$ terms.
As a result, the dominant part of the ideal is a power series in $\epsilon^2$
\begin{equation}
\label{series}
V^*_{dom}=\sum_{k=0}^{\infty} (-)^k\,c_k\,\epsilon^{2k}\ .
\end{equation}
Leaving out for the moment the proof that the signs in the above series alternate we focus on the asymptotic behavior of the coefficients $c_k$
which consist of  products of
$(k+1)$ $V$'s with $2k$ derivatives.
If $V$ diverges as $q^n$ then $c_k\sim A_k\,q^{(k+1)n-2k}$, where $A_k$ are constants that can't be determined from the index alone. From eq.~(\ref{dominant}) it follows that the
$A_k$'s are smaller than one and decrease with $k$.
At $q_c$ we find $c_k\epsilon^{2k}\sim A_k\,\epsilon^{-2n/(n-2)}$, i.e. all the $k$ dependance is in the constants $A_k$.
A similar result follows when $V\sim e^q$.
Therefore at $q_c$ the ideal is proportional to $\sum_k (-)^k A_k$. This is finite since it is smaller than
$\sum_k (-)^k$ which is regularized to $1/2$.

The toy model for the asymptotics of ideals (when all the $A_k$'s are equal to 1) can be solved exactly.
For potentials that diverge as $q^n$ the path integral ideal at a point $\lambda\, q_c$ equals
$W^*_{toy}\sim \lambda^n\epsilon^{-2n/(n-2)}\sum_k (-)^k\lambda^{(n-2)k}=\lambda^n\epsilon^{-2n/(n-2)}/(1+\lambda^{n-2})$.
We find that the alternating sign in eq.~(\ref{series}) plays a crucial role in making this convergent at the critical point ($\lambda=1$).
From $W^*<W^*_{toy}$ it follows that, unlike its truncations at a given $p$ level, the exact path integral ideal remains well behaved even at the critical points.
We conclude that critical points, unboundedness from bellow and truncations of coordinates are all artifacts of the asymptotic expansion procedure
and are not present in the exact ideals.

The only thing that remains is to prove that $V^*_{dom}$ is indeed given in terms of an alternating sum.
To do this we first set $\delta=0$ in eq.~(\ref{integralequation}), and use the fact that the ideal is even in the first argument.
Next, we set $q$ to be very large, a regime where the ideal may be approximated by its dominant term. In this regime
$q\gg y\sim\sqrt{\epsilon}$ so that $V^*_{dom}(q+y/2;\epsilon/2)$ is well approximated by the first two terms in its Taylor expansion about $q$.
Finally, after calculating the remaining Gaussian integral we find
\begin{equation}
\label{approximate}
V^*_{dom}\Big(q;\epsilon\Big)-V^*_{dom}\Big(q;\frac{\epsilon}{2}\Big)\approx -\frac{\epsilon^2}{32}\left[\frac{d}{dq}V^*_{dom}\left(q;\frac{\epsilon}{2}\right)\right]^2\ .
\end{equation}
From this we can directly derive the first two terms of eq.~(\ref{dominant}).
To get all the $V^*_{dom}$ terms correctly it would be necessary to solve its integral equation asymptotically. This is a much simpler
job than was solving eq.~({\ref{integralequation}). Still, eq.~(\ref{approximate}) is good enough for us to extract three important properties of $V^*_{dom}$:
(a) it is given as a power series in $\epsilon^2$,
(b) the terms in this series alternate in sign, (c) and the difference $V^*_{dom}-V$ is proportional to ${V'}^2$.

To conclude, we studied the asymptotic properties of path integral ideals -- expressions that allow us to construct the best discretized effective action for a general quantum theory. Path integral ideals govern convergence flow to the continuum limit allowing us to: (a) systematically accelerate numerical algorithms, (b) investigate the behavior of quantum theories for large times of propagation (low temperatures) using extremely coarse discretizations.
We have shown that ideals are classified according to the degree of divergence of the original potential at spatial infinity, and have isolated dominant terms responsible for the behavior of the ideals of a generic theory for large discrete time steps. We have shown that exact path integral ideals are always well behaved, in contrast to the behavior of some of their $\epsilon^p$ truncations.
%The truncations of the ideal for $p=3,4,7,8,11,12,\ldots$ (and potentials that diverge at spatial infinity faster than $q^2$) are shown to be unbounded from bellow %making it necessary to introduce coordinate cut-offs. The remaining $p$ level truncations are well behaved for all potentials. The fact that exact ideals are well %behaved implies that the unboundedness from bellow is not physical but rather an artifact of asymptotic expansion.

We acknowledge financial support from the Ministry of Science and Environmental Protection of the Republic of Serbia through Projects No. 1486 and No. 1899.
%------------------------------------------------------------------------------

%------------------------------------------------------------------------------

\begin{thebibliography}{24}

\bibitem{feynmanhibbs}
R. P. Feynman and A. R. Hibbs,
\emph{Quantum Mechanics and Path Integrals}
(McGraw-Hill, New York, 1965).
\bibitem{feynman}
R. P. Feynman,
\emph{Statistical Mechanics}
(W. A. Benjamin, New York, 1972).
\bibitem{itzyksonzuberplus}
C. Itzykson and J.-B. Zuber,
\emph{Quantum Field Theory}
(McGraw-Hill, New York, 1980); S. Coleman,
\emph{Aspects of Symmetry}
(Cambridge University Press, 1985).
\bibitem{itzyksondrouffeplus}
C. Itzykson and J.-M. Drouffe,
\emph{Statistical Field Theory}
(Cambridge University Press, 1991); G. Parisi,
\emph{Statistical Field Theory}
(Addison Wesley, New York, 1988).
\bibitem{kleinertplus}
F. Jasch and H. Kleinert,
J. Math. Phys. {\bf 42}, 52 (2001); 
B. Hamprecht and H. Kleinert,
Phys. Rev. D {\bf 68}, 065001 (2003).
\bibitem{barker2}
J. A. Barker and D. Henderson,
Rev. Mod. Phys. {\bf 48}, 587 (1976); J. A. Barker,
J. Chem. Phys. {\bf 70}, 2914 (1979).
\bibitem{ceperley2}
E. L. Pollock and D. M. Ceperley,
Phys. Rev. B {\bf 30}, 2555 (1984);
D. M. Ceperley,
Rev. Mod. Phys. {\bf 67}, 279 (1995).
\bibitem{kleinert}
H. Kleinert,
\emph{Path Integrals in Quantum Mechanics, Sta\-tistics, Polymer Physics, and Financial Markets}
(World Scientific, 2004).
\bibitem{kroegerplus}
H. Kr\"oger,
Phys. Rep. {\bf 323}, 81 (2000);
A. Bogojevi\'c, A. Bala\v z and A. Beli\'c,
Phys. Lett. A {\bf 345}, 258 (2005).
\bibitem{prlprb}
A. Bogojevi\'c, A. Bala\v z and A. Beli\'c,
Phys. Rev. Lett. {\bf 94}, 180403 (2005);
Phys. Rev. B {\bf 72}, 064302 (2005).
\bibitem{pla05a}
A. Bogojevi\'c, A. Bala\v z and A. Beli\'c,
Phys. Lett. A {\bf 344}, 84 (2005).
\end{thebibliography}
\end{document}